\documentclass[10pt,notitlepage,preprintnumbers,nofootinbib,showpacs]{revtex4-1}

\usepackage{color,graphicx} \usepackage{ifpdf}
\usepackage{amsmath} \usepackage{amssymb} \usepackage{bm}

\usepackage{hhline}
\usepackage{color}
\definecolor{nicered}{rgb}{0.7,0.1,0.1}
\definecolor{nicegreen}{rgb}{0.1,0.5,0.1}

\usepackage[english]{babel}
\usepackage{slashed}
\usepackage{multirow}

\pdfoutput=1

\usepackage{braket}
\usepackage{slashed}

\newcommand {\E}[1]{\times 10^{#1}}	
\newcommand {\e}[1]{\mathrm{~#1}}       
\newcommand{\mc}[1]{\mathcal{#1}}

\newcommand{\mrm}[1]{\mathrm{#1}}
\newcommand{\re}[0]{\mrm{Re}}

\newcommand{\bea}{\begin{eqnarray}}
\newcommand{\eea}{\end{eqnarray}}

\newcommand{\br}[0]{\mc{B}}

\definecolor{Red}{rgb}{1.,0.,0.}

\definecolor{Green}{rgb}{0.2,.7,0.2}

\usepackage{mciteplus}

\usepackage{hyperref}
\hypersetup{colorlinks,citecolor=nicegreen,linkcolor=nicered}
\hypersetup{colorlinks=true}

\begin{document}

\author{Svjetlana Fajfer} \email[Electronic
address:]{svjetlana.fajfer@ijs.si} 
\affiliation{Department of Physics,
  University of Ljubljana, Jadranska 19, 1000 Ljubljana, Slovenia}
\affiliation{J. Stefan Institute, Jamova 39, P. O. Box 3000, 1001
  Ljubljana, Slovenia}

\author{Nejc Ko\v snik} 
\email[Electronic address:]{nejc.kosnik@ijs.si}

\affiliation{J. Stefan Institute, Jamova 39, P. O. Box 3000, 1001 Ljubljana, Slovenia}

\title{Vector leptoquark resolution of $R_K$ and $R_{D^{(*)}}$ puzzles}

\date{\today}

\begin{abstract}
  We propose that three recent anomalies in $B$ meson decays,
  $R_{D^{(*)}}$, $R_K$, and $P_5'$, might be explained by a single
  vector leptoquark weak triplet state.  The constraints on the
  parameter space are obtained by considering $t \to b \tau^+ \nu$
  data, lepton flavor universality tests in the kaon sector, bounds on $B \to K^{(*)}
  \bar \nu \nu$, bound on
  the lepton flavor violating decay $B \to K \mu \tau$, and
  measurements of $b \to c \mu^- \bar\nu$ decays.  The presence of such vector
  leptoquark could be exposed in precise measurements of $t \to b \tau
  \nu$ and $B \to K^{(*)}
  \bar \nu \nu$ decays. The model also predicts approximate equality of lepton
  flavor universality ratios $R_{K^*}$, $R_{K}$, and suppressed
  branching fraction of $B_s \to \mu^+ \mu^-$.
\end{abstract}

\maketitle

\section{Introduction}
Although LHC has not found yet any particles not present in the
Standard Model~(SM), low-energy precision experiments in $B$ physics
pointed out a few puzzling results.  Namely, we are witnessing
persistent indications of disagreement with the SM prediction of
lepton flavor universality~(LFU) ratio in the $\tau/\mu$ and/or $\tau/e$
sector. In the case of ratio
$R_{D^{(*)}} = \tfrac{\Gamma(B \to D^{(*)} \tau^- \bar\nu)}{\Gamma(B \to
  D^{(*)} \ell^- \bar\nu)}$~\cite{Lees:2012xj,
  Lees:2013uzd, Huschle:2015rga, Adachi:2009qg, Bozek:2010xy,
  Aaij:2015yra}, 
the deviation from the SM is at 3.5$\sigma$ level~\cite{Bauer:2015knc}
and has attracted a lot of attention recently~\cite{Crivellin:2014zpa,Bhattacharya:2014wla,Bhattacharya:2015ida,Hati:2015awg,Sakaki:2014sea}. Since the
denominator of these ratios are the well measured decay rates with
light leptons in the final states, $\ell = e,\mu$, the most obvious
interpretation of $R_{D^{(*)}}$ results are in terms of new physics
affecting semileptonic $b \to c \tau^- \bar\nu$
processes~\cite{Fajfer:2012jt}.

The second group of observables, testing rare neutral current
processes with flavor structure $(\bar s b)(\mu^+ \mu^-)$ also
indicate anomalous behaviour~\cite{Altmannshofer:2013foa, Altmannshofer:2014cfa, Hiller:2014yaa,Glashow:2014iga,
Gripaios:2014tna, Ghosh:2014awa, Crivellin:2015mga,
Crivellin:2015lwa, Sierra:2015fma, Varzielas:2015iva,
Crivellin:2015era, Celis:2015ara, Belanger:2015nma, Pas:2015hca}. Decay
$B \to K^{*} \mu^+ \mu^-$ deviates from the SM in the by-now-famous
$P_5'$ angular observable at the confidence level of above
$3\sigma$~\cite{DescotesGenon:2012zf,Descotes-Genon:2013wba,Altmannshofer:2014rta}. If interpreted in terms of
new physics~(NP), all analyses point to modifications of the leptonic
vector current, which is also subject to large uncertainties due to
nonlocal QCD effects. However, several studies have shown that even
with generous errors assigned to QCD systematic effects, the anomaly
is not washed away~\cite{Jager:2014rwa}. Furthermore, the sizable
violation of LFU in the ratio
$R_K = \tfrac{\Gamma(B \to K \mu \mu)}{\Gamma(B \to K ee)}$ in the
dilepton invariant mass bin 1 GeV$^2 \leq q^2 \leq$ 6 GeV$^2$ has
been established at $2.6\sigma$ level. This ratio, being largely free of
theoretical uncertainties and experimental systematics, deviates in
the muon channel consistently with the deviation in
$B \to K^* \mu^+ \mu^-$.  Strikingly enough, all these disagreements were
observed in the $B$ meson decays to the leptons of the second and third
generation. As pointed out in~\cite{Fajfer:2012jt} the lepton flavour
universality has been tested at percent level and is, in the case of pion
and kaon, in excellent agreement with the SM predictions.  It has been
already suggested that scalar leptoquark might account for this
anomalous behaviour in the recent literature~\cite{Hiller:2014yaa,Sahoo:2015wya,
Becirevic:2015asa, Freytsis:2015qca, Bauer:2015knc, Gripaios:2014tna, Datta:2013kja}.

Many models of NP~\cite{Altmannshofer:2013foa, Datta:2013kja, Hiller:2014yaa,
Crivellin:2014zpa,Glashow:2014iga,Bhattacharya:2014wla,Gripaios:2014tna,
Ghosh:2014awa, Crivellin:2015mga, Crivellin:2015lwa, Sierra:2015fma,
Varzielas:2015iva, Crivellin:2015era, Celis:2015ara,Freytsis:2015qca} have been employed to
explain either $R_K$ and $P_5^\prime$ anomalies or
$R_{D^{(*)}}$. It was suggested in Ref.~\cite{Bhattacharya:2014wla} that $R_K$ and
$P_5^\prime$ can be explained if NP couples only to the third
generation of quarks and leptons. Similarly, the authors
of~\cite{Calibbi:2015kma} suggested that both $R_{D^{(*)}}$ and
$R_K$ anomalies can be correlated if the effective four-fermion
semileptonic operators consist of left-handed doublets. The model
of~\cite{Greljo:2015mma} proposed existence of an additional weak
bosonic triplet and falls in the category of weak doublet fermions coupling to
the weak triplet bosons, which then can explain all three $B$ meson
anomalies. Among the NP proposals a number of them suggest that one scalar leptoquark accounts for either $R_D^{(*)}$ or $R_K$ anomalies.  
However, in the recent paper~\cite{Bauer:2015knc} both deviations were
addressed by a single scalar leptoquark with quantum numbers
$(3,1,-1/3)$ in such a way that $R_{D^{(*)}}$  anomalies are explained
at the tree level, while $R_K$ receives contributions at loop level. 
This scalar leptoquark unfortunately can couple to a diquark state too
and therefore it potentially leads to proton decay. One may impose
that this dangerous coupling vanishes, but such a scenario is not
easily realised within Grand Unified Theories. 

In this paper, we extend the SM by a vector $SU(2)$ triplet leptoquark, which 
accomplishes both of the above requirements by generating purely left-handed currents with quarks and leptons. 
Furthermore, the triplet
nature of the state connects the above mentioned anomalies with the
rare decay modes of $B$ mesons to a final states with neutrinos, and
various charged lepton flavor violating decay modes.
The considered state has no couplings to diquarks and has therefore definite
baryon and lepton numbers and does not mediate proton decay. In~\cite{Calibbi:2015kma} the
same leptoquark state has been considered in a more restricted
scenario with couplings to the third generation fermions in the weak basis.

The outline of this paper is the following: In Sec. II we describe how to
accommodate  $R_{D^{(*)}}$ and $R_K$ within the scenario where vector triplet
leptoquark mediates quark and lepton interactions. Sec. III
discusses current constraints on the model and further experimental signatures of this
model, while in the last Section we present conclusions.

\section{Signals}
The vector multiplet $U_3^\mu$ that transforms under the SM gauge group as
$(3,3,2/3)$
couples to a leptoquark current with $V-A$ structure:
\begin{equation}
  \label{eq:LQLag}
  \mc{L}_{U_3} = g_{ij} \bar Q_i \gamma^\mu \,\tau^A  U^A_{3\mu}\, L_j + \mrm{h.c.}.
\end{equation}
Here $\tau^A, A = 1,2,3$ are the Pauli matrices in the $SU(2)_L$
space whereas $i,j = 1,2,3$ count generations of the left-handed lepton
and quark doublets, $L$ and $Q$, respectively. The couplings $g_{ij}$
are in general complex parameters, while for the sake of simplicity we
will restrict our attention to the case where they are real. The absence of any other
term at mass dimension 4 of the operators ensures the conservation of baryon and lepton
numbers and this allows the leptoquark $U_{3}$ to be close to the TeV scale
without destabilizing the proton. The interaction Lagrangian~\eqref{eq:LQLag} is written in
the mass basis with $g_{ij}$ entries defined as the couplings between
the $Q=2/3$ component of the triplet, $U_{3\mu}^{(2/3)}$, to $\bar d_{Li}$ and
$\ell_{Lj}$. Remaining three types of vertices to eigencharge states
$U_{3\mu}^{(2/3)}$, $U_{3\mu}^{(5/3)}$, and $U_{3\mu}^{(-1/3)}$ are then obtained by
rotating the $g$ matrix, where necessary, with the
Cabibbo-Kobayashi-Maskawa (CKM) matrix $\mc{V}$ from the left
or with the Pontecorvo-Maki-Nakagawa-Sakata (PMNS) matrix $\mc{U}$ from the right:
\begin{equation}
  \label{eq:LQLagComp}
  \begin{split}
      \mc{L}_{U_3} =\,\, &U^{(2/3)}_{3\mu} \, \Big[ (\mc{V} g
        \mc{U})_{ij}\, \bar u_i \gamma^\mu P_L \nu_j  -
        g_{ij} \,\bar d_i \gamma^\mu P_L \ell_j \Big]\\
 &+U^{(5/3)}_{3\mu}\, (\sqrt{2} \mc{V} g)_{ij}\, \bar u_i \gamma^\mu P_L
 \ell_j \\
  &+U^{(-1/3)}_{3\mu}\, (\sqrt{2} g \mc{U})_{ij}\, \bar d_i \gamma^\mu P_L
 \nu_j+\mrm{h.c.}.
  \end{split}
\end{equation}
If ultraviolet origin of the $U_3^\mu$ LQ is a gauge boson field of some higher
symmetry group (e.g. Grand Unified Theory), then the coupling matrix $g$
in the mass basis should be unitary. Furthermore, in such theories
the ability to choose gauge and the presence of additional Goldstone degrees of freedom
would ensure renormalizability, in contrast to the effective theory
of Eq.~\eqref{eq:LQLag}. In this work we limit ourselves to the
tree-level constraint for which the details of the underlying ultraviolet
completion are irrelevant.

The $b \to s \mu^+ \mu^-$ processes are affected by the product $g_{b
  \mu}^* g_{s \mu}$ whereas the crucial parameter for $b \to c \tau^- \bar \nu$
is $g_{b \tau}$. We do not insist on a particular flavor structure of
the matrix $g$ but note that the explanation of 
the LFU puzzles in the neutral and charged currents involves
parameters $g_{s\mu}$,
$g_{b \mu}$, and $g_{b \tau}$, which will be our tunable flavor
parameters of the model. We assume the remaining
elements $g_{ij}$ are negligibly small:
\begin{equation}
\label{eq:texture}
g =
\begin{pmatrix}
  0 & 0 & 0\\
  0 & g_{s \mu} & 0\\
  0 & g_{b \mu} & g_{b \tau}
\end{pmatrix},
\qquad
\mc{V} g =
\begin{pmatrix}
  0 & \mc{V}_{us} g_{s\mu} + \mc{V}_{ub} g_{b\mu} & \mc{V}_{ub} g_{b\tau}\\
  0 & \mc{V}_{cs} g_{s\mu} + \mc{V}_{cb} g_{b\mu} & \mc{V}_{cb} g_{b\tau}\\
  0 & \mc{V}_{ts} g_{s\mu} + \mc{V}_{tb} g_{b\mu} & \mc{V}_{tb} g_{b\tau}
\end{pmatrix}.
\end{equation}
The rotated matrix $\mc{V} g$ determines the couplings of the LQ to
the up-type quarks among which we also have a $U_{3\mu}^{(2/3)}$ coupling
to $\bar c \nu$, required to explain $R_{D^{(*)}}$. 

The leptoquark $U_{3}$ implements a combination of Wilson
coefficients in the $b \to s \mu^+ \mu^-$ effective
Lagrangian~\cite{Gripaios:2014tna, Kosnik:2012dj},
\begin{equation}
\label{eq:C9}
C_9 = -C_{10} =  \frac{\pi}{\mc{V}_{tb} \mc{V}_{ts}^*
  \alpha}\, g_{b \mu}^* g_{s \mu}\,\frac{v^2}{M_U^2},
\end{equation}
which has been shown to significantly improve the global fit of the
$b \to s \mu^+ \mu^-$ observables with the $1\sigma$ preferred region
$C_9 \in [-0.81,-0.50]$~\cite{Descotes-Genon:2015uva}, see
also~\cite{Alonso:2015sja}. Here $v=246\e{GeV}$ is the electroweak
vacuum expectation value. In this case we find
\begin{equation}
g_{b \mu}^* g_{s \mu} \in [0.7, 1.3] \E{-3} \,\left(M_U/\mrm{TeV}  \right)^2.
\end{equation}
Note that the effective coupling \eqref{eq:C9} also brings the LFU
observable $R_K$ in agreement with the experimental
value~\cite{Descotes-Genon:2015uva}.

On the other hand, the correction to the semileptonic decays $b \to c \tau^- \bar \nu$
also proceeds via exchange of the $U^{(2/3)}_{3\mu}$ state. The
effective semileptonic Lagrangian in  the SM complemented by the LQ correction is:
\begin{equation}
  \label{eq:LagSL}
  \begin{split}
  \mc{L}_\mrm{SL} &= -\left[\frac{4G_F}{\sqrt{2}} \mc{V}_{cb} \mc{U}_{\tau
    i} +
    \frac{g_{b\tau}^* (\mc{V} g \mc{U})_{ci}}{M_U^2}\right] (\bar c
    \gamma^\mu P_L b)(\bar \tau \gamma_\mu P_L \nu_i)  + \mrm{h.c.}
  \end{split}
\end{equation}
The second term shifts the effective value of $|\mc{V}_{cb}|^2$
as measured in semitauonic decays summed over all neutrino species in the
final state:
\begin{equation}
\label{eq:Vcbtau}
\left|\mc{V}_{cb}^{(\tau)}\right|^2 \simeq |\mc{V}_{cb}|^2 \left[1+ \frac{v^2}{M_U^2}
  \re\left(\frac{g_{b\tau}^* (\mc{V} g)_{c\tau}}{\mc{V}_{cb}}\right)\right].
\end{equation}
The above expression contains the interference term with the SM
amplitude while the pure LQ contribution is rendered negligible
compared to the interference term
by an additional factor $v^2/M_U^2$.
In the same manner the semimuonic decay widths $b \to c \mu^- \bar \nu$
are proportional to $|\mc{V}_{cb}^{(\mu)}|^2$ that is given by an
analogous expression to Eq.~\eqref{eq:Vcbtau}.  From the fit to the
measured ratio $R_{D^{(*)}}$ done in Ref.~\cite{Freytsis:2015qca} we
learn that at $1\sigma$ we have the following constraint:
\begin{equation}
\re\left[g_{b\tau}^*    (\mc{V} g)_{c\tau} - g_{b\mu}^*    (\mc{V}
  g)_{c\mu} \right]= (0.18 \pm 0.04)\, \left(M_U/\mrm{TeV}  \right)^2.
\end{equation}
We are allowing for LQ modifications to take place for both $\ell =
\mu,\tau$ in $b \to
c \ell^- \bar \nu$.

In summary, the data on $b \to s \mu^+ \mu^-$ and
$R_{D^{(*)}}$ points to a region in parameter space where
\begin{equation}
  \begin{split}
  g_{b\mu} g_{s\mu}& \approx 10^{-3},\\
\mc{V}_{cb} (g_{b\tau}^2-g_{b\mu}^2) -  g_{b\mu} g_{s\mu} &\approx 0.18,
\end{split}
\end{equation}
is satisfied, if $M_U = 1\e{TeV}$. From the first equation we learn that,
once we impose perturbativity condition ($|g_{s\mu},g_{b\mu},
g_{b\tau}| < \sqrt{4\pi}$), that both $|g_{s\mu}|$ and $|g_{b\mu}|$ are
also bounded from below, $|g_{s\mu}|, |g_{b\mu}| \gtrsim 3\E{-4}$. The
second equation can be simplified to
\begin{equation}
g_{b\tau}^2 - g_{b\mu}^2 \approx 4.4,
\end{equation}
which indicates $|g_{b\tau}| \gtrsim 2$.

\section{Additional constraints}
\subsection{LFU in the kaon sector}
Potentially very severe constraints are the measurements of $|\mc{V}_{us}|$
in kaon muonic decays due to $U_{3\mu}$ contributions in
$s \to u \mu^- \bar \nu$ but not in $s \to u e^- \bar \nu$, since first
generation charged leptons are not affected by the studied LQ at tree
level. Effects of this type are exposed by the lepton
flavor universality ratios between decays involving the kaon and
different charged leptons:
\begin{equation}
R_{e/\mu}^K  = \frac{\Gamma(K^- \to e^- \bar \nu)}{\Gamma(K^- \to \mu^-
  \bar \nu)}, \qquad
R_{\tau/\mu}^K  = \frac{\Gamma(\tau^-\to K^- \nu)}{\Gamma(K^- \to \mu^- \bar \nu)}.
\end{equation}
Note that the value of $|\mc{V}_{us}|$ obtained from the global CKM fits
relies on the data on semielectronic decays (cf. experimental inputs
to $\mc{V}_{us}$ of the CKMfitter results~\cite{Charles:2004jd} prepared
for the EPS 2015 conference) that are not subject to the
leptoquark amplitudes. The SM value of $|\mc{V}_{us}|$ is thus not a
relevant constraint on the leptoquark couplings. The measured value of
$R^K_{e/\mu}$ is due to the NA62 experiment~\cite{Lazzeroni:2012cx}
while the SM prediction has been calculated with negligible
uncertainty~\cite{Cirigliano:2007xi} and is in good agreement with
the experimental result:
\begin{equation}
\label{eq:Remu}
  R_\mrm{e/\mu}^{K(\mrm{exp})} = (2.488\pm 0.010)\E{-5},\qquad
  R_\mrm{e/\mu}^{K(\mrm{SM})} =(2.477\pm 0.001)\E{-5}.
\end{equation}
In the $\tau/\mu$ sector, the SM prediction and the
value obtained from the measured branching fractions~\cite{Agashe:2014kda} agree as well:
\begin{equation}
\label{eq:Rtaumu}
  R_\mrm{\tau/\mu}^{K(\mrm{exp})} = (1.101\pm 0.016)\E{-2},\qquad
  R_\mrm{\tau/\mu}^{K(\mrm{SM})} = (1.1162 \pm 0.00026)\E{-2}.
\end{equation}
From the Lagrangian~\eqref{eq:LQLagComp} and couplings~\eqref{eq:texture} one
can derive the LQ modification of $\mc{V}_{us}$ as measured in $s \to u
\mu^- \bar \nu$ decay:
\begin{equation}
  \begin{split}
\mc{V}_{us}^{(\mu)} &= \mc{V}_{us} \left[1 + \frac{v^2}{2 M_U^2}
  \re\left(\frac{g_{s\mu}^*  (\mc{V} g )_{u \mu}}{\mc{V}_{us}}\right)\right]\\
&\equiv \mc{V}_{us} \left[1 + \delta_{us}^{(\mu)} \right].
  \end{split}
\end{equation}
Again, we have neglected the pure LQ terms which are
proportional to $v^4/M_U^4$.
The presence of LQ modifies both LFU ratios $R_{e/\mu}^{K}$,
$R_{\tau/\mu}^{K}$ by a common factor
\begin{equation}
    R_{\ell/\mu}^{K\mrm{(SM)}}  \to  R_\mrm{\ell/\mu}^{K(\mrm{SM})}
    \left[1 - 2\delta_{us}^{(\mu)} \right], \qquad \ell = e, \tau.
\end{equation}
We determine $\delta_{us}^{(\mu)} = (-2.2\pm 2.2)\E{-3}$ and
$\delta_{us}^{(\mu)} = (6.7\pm 7.1)\E{-3}$ using the
$e/\mu$~\eqref{eq:Remu} and $\tau/\mu$~\eqref{eq:Rtaumu} LFU ratios,
respectively. Combining the two determinations of
$\delta_{us}^{(\mu})$ results in average value
$\delta_{us}^{(\mu)} = (-1.4\pm 2.1)\E{-3}$ and allows to
put constraint on the LQ couplings:
\begin{equation}
\re\left(|g_{s\mu}|^2 + \frac{\mc{V}_{ub}}{\mc{V}_{us}} g_{s\mu}^*
  g_{b\mu}\right) = (-4.6\pm 6.9)\E{-2} (M_U/\mrm{TeV})^2.
\end{equation}

\subsection{Semitauonic top decays}
The eigencharge state $U^{(2/3)}_{3\mu}$ can have large effects also in semileptonic
decays of the top quarks, in particular in the decay mode $t \to b \tau^+
\nu$ being a purely third-generation transition. The correction
to the tau-specific CKM element $\mc{V}_{tb}$ reads
\begin{equation}
\label{eq:tbtau}
  \mc{V}_{tb}^{(\tau)} = \mc{V}_{tb} \left[1+\delta_{tb}^{(\tau)}\right],\qquad
  \delta_{tb}^{(\tau)} = \frac{v^2}{2 M_U^2}\,\re\left(\frac{g_{b\tau}^*
    (\mc{V} g)_{t\tau}}{\mc{V}_{tb}}\right).
\end{equation}
The correction $\delta_{tb}^{(\tau)}$ should be smaller than the
relative error on $\mc{V}_{tb}$ as measured in decay $\mc{B}(t \to b \tau^+
\nu) = 0.096\pm 0.028$ by the CDF
collaboration~\cite{Aaltonen:2014hua}:
\begin{equation}
\frac{v^2}{M_U^2}\,\re\left(\frac{g_{b\tau}^*
    (\mc{V} g)_{t\tau}}{\mc{V}_{tb}}\right) < 0.29.
\end{equation}
This bound can be interpreted as
\begin{equation}
|g_{b\tau}| < 2.2\, (M_U/\mrm{TeV}).  
\end{equation}
Recent analysis of the top
decays in the $t\bar t$ production channel already probes $\mc{V}_{tb}$ in
semitauonic decays of the top quark with competitive
precision~\cite{Aad:2015dya, Jernej}.

\subsection{$b\to c \mu^- \bar\nu$ decay}
For the rate of the semimuonic decays we are not aware,
to our best knowledge, of an experimental measurement of $B \to D \ell^-
\bar\nu$ quoting separate lepton-specific rates for $\ell = e$ and $\ell =
\mu$. From the data on the semileptonic decays $b \to c \ell^- \bar
\nu$ the average of inclusive and exclusive determinations is $|\mc{V}_{cb}|_\mrm{exp.} =
(41.00 \pm 1.07)\E{-3}$, a value reported by the HFAG~\cite{Amhis:2014hma} and used by
the CKMfitter group. On the other hand, CKMfitter 
performed a fit without using $|\mc{V}_{cb}|_\mrm{exp.}$ as input and the
preliminary result is
then $|\mc{V}_{cb}|_\mrm{indirect} = (42.99^{+0.36}_{-1.41})\E{-3}$~\cite{Charles:2004jd}. The 
difference between experimental and indirect determination of $\mc{V}_{cb}$
can then be assigned to the leptoquark contribution:
\begin{equation}
\label{eq:semimuonic}  
\begin{split}
|\mc{V}_{cb}|_\mrm{exp.} - |\mc{V}_{cb}|_\mrm{indirect} &= (-2.0^{+1.7}_{-1.2})\E{-3}\\
  &=\frac{v^2}{2M_U^2} |\mc{V}_{cb}| \re\left(\frac{g_{b\mu}^*
    (\mc{V} g)_{c\mu}}{\mc{V}_{cb}}\right).
\end{split}
\end{equation}
The ensuing constraint is
\begin{equation}
|\mc{V}_{cb}| \re\left(\frac{g_{b\mu}^*    (\mc{V} g)_{c\mu}}{\mc{V}_{cb}}\right)\in [-0.1,
  -0.01]\E{-3} \left(M_U/\mrm{TeV}  \right)^2.  
\end{equation}
Notice that the considered leptoquark does not affect the
semielectronic decays, and that the entire effect originates from
semimuonic decays in our model. Although the presented bound includes 
intrinsic pollution from the semielectronic events, in lack of better
constraint, we apply it as a bound on
the LQ modification of semimuonic decays. It would be indeed very useful to have experimental results on
the semileptonic rates for different leptons in the final states.

\subsection{$B \to K \mu  \tau$ decay}
The observables that probe the LQ couplings with the $b$ quark and
violate lepton flavor are, at tree level, $B^- \to K^- \mu^+ \tau^-$
and decays of bottomonium to $\tau \mu$. The branching ratio of the
latter process is constrained at the level of $10^{-6}$ but taking
into account large decay widths of bottomonia states, these bounds are
not competitive with the bound
$\br(B^- \to K^- \mu^+ \tau^-) < 2.8\E{-5}$ at 90\%
CL~\cite{Lees:2012zz}. 
We can estimate the decay width by adapting the bound from
the very same process analysed in the
case of scalar leptoquark in the representation $(\bar 3,1,4/3)$~\cite{Dorsner:2011ai}:
\begin{equation}
|g_{b \tau} g_{s\mu}| \lesssim 0.09 (M_U/\mrm{TeV})^2.  
\end{equation}

\subsection{$B \to K^{(*)} \nu \bar \nu$ decay}
The $B \to K^{(*)} \nu \bar\nu$ probes lepton flavor conserving as
well as lepton flavor violating combination of the LQ couplings. Using the
notation of Refs.~\cite{Altmannshofer:2009ma,Buras:2014fpa} and
extended in \cite{Becirevic:2015asa} to account for lepton flavor
violation, we employ the effective Lagrangian
\begin{equation}
\mc{L}_\mrm{eff}^{b\to s \bar \nu \nu} = \frac{G_F \alpha }{\pi \sqrt{2}} \mc{V}_{tb} \mc{V}_{ts}^*
C_L^{ij} (\bar s \gamma_\mu P_L b)(\bar \nu_i \gamma^\mu (1-\gamma_5) \nu_j).
\end{equation}
The effect of the $U_3$ leptoquark has been already studied in
\cite{Buras:2014fpa}. In the SM we have, for each pair of neutrinos,
$C_L^{\mrm{SM},ij} = C_L^\mrm{SM}\delta_{ij}$, where $C_L^\mrm{SM} =
-6.38 \pm 0.06$~\cite{Altmannshofer:2009ma}. On the other hand, the
vector LQ generates $C_L^{\mrm{LQ},ij} = 2 \pi (g U)_{bi}^* (g U)_{sj}
v^2/(\alpha \mc{V}_{tb} \mc{V}_{ts}^* M_U^2)$. 
The branching ratios of  $B \to K^{(*)} \bar \nu \nu$ --- defined as a
sum over branching fraction for each combination of neutrino species in the final
state --- get modified by the same factor for both $K$ and $K^*$ decay
modes~\cite{Buras:2014fpa}:
\begin{equation}
\label{eq:nunu-factor}
1+\frac{4\pi v^2}{3  \alpha \mc{V}_{tb} \mc{V}_{ts}^* M_U^2
  C_L^\mrm{SM}} \re(g_{s\mu} g_{b\mu}^*) + \frac{1}{3|C_L^\mrm{SM}|^2}
\left(\frac{2\pi v^2}{\alpha \mc{V}_{tb} \mc{V}_{ts}^* M_U^2}
\right)^2 |g_{s\mu}|^2 \left(|g_{b\mu}|^2 + |g_{b\tau}|^2 \right).
\end{equation}
The LQ prediction of $\mrm{Br}(B \to K^{(*)}\nu \bar\nu)$ is thus obtained by
rescaling the SM prediction, e.g.
$\mrm{Br}(B^+ \to K^+ \nu \bar\nu) = (4.0\pm 0.5)\E{-6}$, by
factor~\eqref{eq:nunu-factor}. Notice that due to the large coupling
$g_{b\tau}$ the most important contribution is the LFV contribution of
the last term in \eqref{eq:nunu-factor}. Imposing the $90\%$
C.L. experimental bound $\mrm{Br}(B^+ \to K^+ \nu \bar\nu)  <
1.6\E{-5}$~\cite{Lees:2013kla} then constrains same coupling combination as the LFV decay $B
\to K \mu \tau$.


\subsection{Fitting the couplings}
\begin{figure}[!h]
    \begin{center}\hspace{2cm}\includegraphics[width=0.95\textwidth]{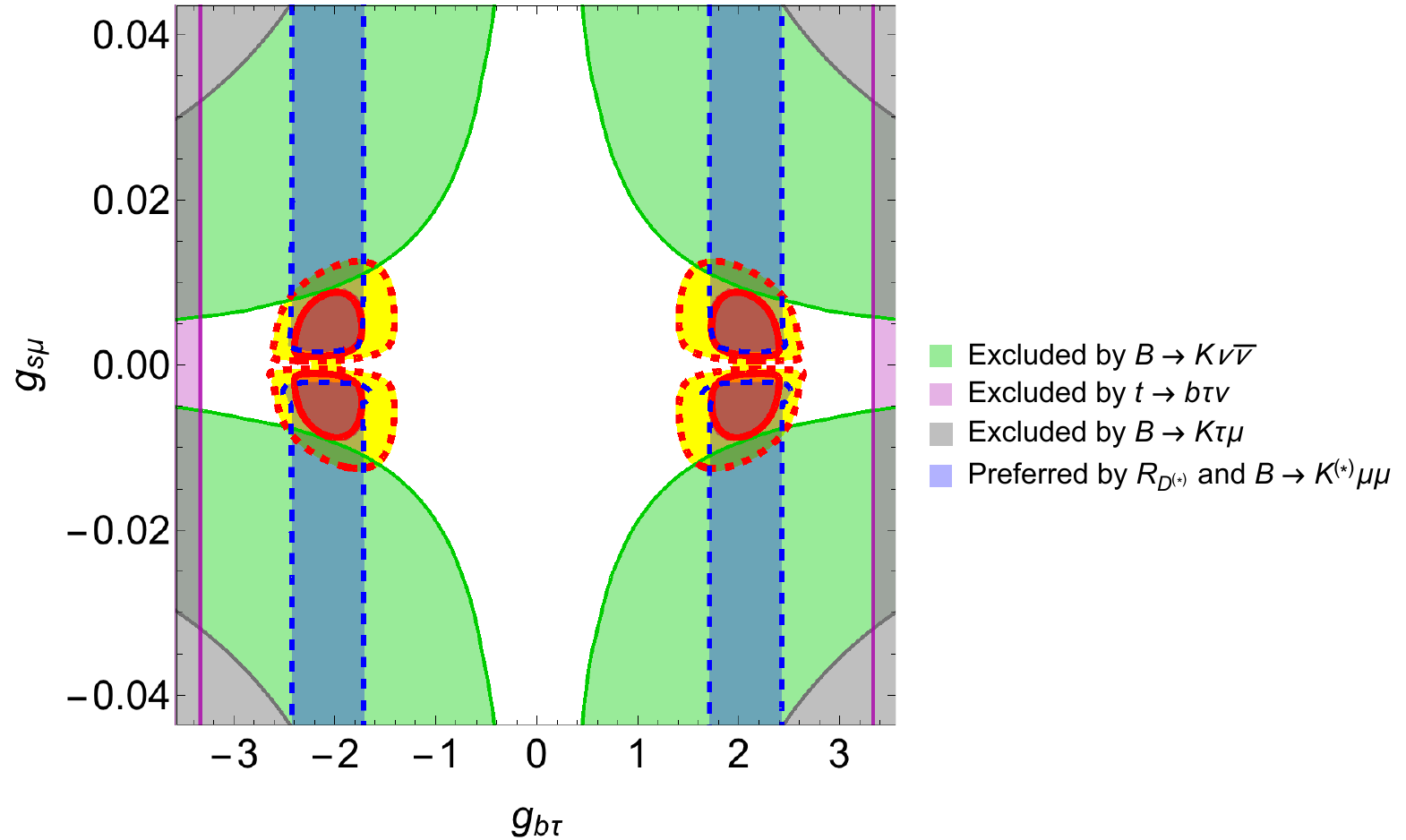}
\end{center}
  \caption{Constraints of real parameters $g_{s\mu}$
    and $g_{b\tau}$ in units $M_U/\mrm{TeV}$. The fitted regions are outlined
    in red ($1\sigma$) and red dashed ($2\sigma$). The region
    preferred by $R_{D^{(*)}}$ and $b \to s \mu^+ \mu^-$ data is
    enclosed by blue dashed contour.}
  \label{fig:ybmuysmu}
\end{figure}

In Fig.~\ref{fig:ybmuysmu} we show the effect of the constraints
projected onto $g_{s\mu}$-$g_{b\tau}$ space; $g_{b\mu}$ is free
parameter of the fit. The best fit point with all the constraints and
signals included is obtained at $\chi^2 \simeq 3$ and is much favoured
over the SM situation. Clearly there is preference for large
$g_{b\tau}$ to correct the large SM tree-level effect in
$b \to c \tau^- \bar\nu$. On the other hand, $g_{s \mu}$ is two
orders of magnitude smaller, and is responsible, together with
moderately large $g_{b\mu}$ ($0.1 \lesssim |g_{b\mu}| \lesssim 1$, not shown in Fig.~\ref{fig:ybmuysmu}),
for the correction of the 1-loop SM effect in $b\to s \mu^+ \mu^-$.

\subsection{Further experimental signatures}
Consequences of the vector LQ for rare charm decays can be extracted
from the couplings of the $U^{(5/3)}_3$ in
Eq.~\eqref{eq:LQLagComp}. One can easily derive the contribution to
the $c \to u \mu^+ \mu^-$ effective Lagrangian. Following notation of
Ref.~\cite{Fajfer:2015mia}, one can easily find that there is
contribution to $C_{9,10}^{(\bar u c)}$ Wilson coefficients:
\begin{equation}
C_9^{(\bar u c)}= -C_{10}^{(\bar u c)} = \frac{2 \pi (\mc{V} g)_{u\mu}
    (\mc{V}g)^*_{c\mu} }{\mc{V}_{ub}\mc{V}_{cb}^*\alpha} \frac{v^2}{M_U^2}.
\label{charm-rare}
\end{equation}
We find $|\tilde C_9| \equiv| C_9^{(\bar u c)}/(\mc{V}_{ub}\mc{V}_{cb}^*)|\lesssim
0.05$, an order of magnitude below the currently allowed bound $|\tilde C_{9}| \leq 0.63$~\cite{Fajfer:2015mia}.

One of the most sensitive channels to test this model is the decay
$t \to b \tau^+ \nu$ which was already used to constrain the
couplings. The largest coupling $g_{b\tau}$ which drives this top decay
is large, $|g_{b\tau}| \sim 2$, and according to Eq.~\eqref{eq:tbtau} 
it increases the decay rate by $20\%$.

In addition, the $U_3$ leptoquark contributes to
$R_{K^*} = \Gamma( B\to K^* \mu^+ \mu^-)/ \Gamma( B\to K^* e^+ e^-)$.
As already discussed in~\cite{Hiller:2014ula}, in scenarios with
left-handed currents the two LFU ratios, $R_{K^*}$ and $R_{K}$, are
predicted to be approximately equal, where the only difference between
them originates from the small quadratic term of the LQ
amplitude. Future LHCb measurements of $R_{K^*}$ will definitely help
in differentiation between different models. Another immediate
consequence of positive LQ contribution to the $C_{10}$, ranging from
$0.4$ to $0.8$ at 1$\sigma$ CL, is destructive interference with the
negative $C_{10}^\mrm{SM}$, which results in $20-35\%$ smaller
branching fraction compared to the SM face value for the time
integrated branching fraction $\mrm{Br}(B_s\to \mu^+ \mu^-)_\mrm{SM} = (3.65 \pm 0.23)\E{-9}$~\cite{Bobeth:2013uxa}.

\section{Conclusions}
We propose that the simple extension of the SM by vector leptoquark
that is a weak triplet can simultaneously explain all three recent $B$
physics anomalies: $R_{D^{(*)}}$, $R_K$, and the $P_5'$ puzzle in $B
\to K^{*} \mu^+ \mu^-$.  The considered triplet LQ contains massive vector states with
electric charges $5/3$, $2/3$ and $-1/3$. The coupling of the charge
$2/3$ state with the second and third generation of down quarks and
charged leptons introduces, via CKM and PMNS mixing, coupling of the
$2/3$ state to the up-type quarks and neutrinos, charge $-1/3$ state
to the down-type quarks and neutrinos, and couplings of charge $5/3$
state to up-type quarks and charged leptons. Our model is constrained
by a number of tree level processes in addition to the $B$ physics
anomalies: tests of lepton flavor universality in $K$ physics, bounds
on decay $B \to K^{(*)} \nu\bar \nu$, semileptonic top decays
$t \to b \tau^+ \nu$, $b\to c \ell^- \bar\nu$ transition, and lepton
flavor violating decay $B \to K \mu \tau$.  The considered vector
leptoquark also affects $c \to u \mu^+ \mu^-$ with the most stringent
constraint coming from $D^0 \to \mu^+ \mu^-$ decay branching fraction
as noticed in~\cite{Fajfer:2015mia}. However, our prediction for the
appropriate Wilson coefficients $C_{9,10}$ turned out to be much
smaller than the ones allowed by the experimental data as discussed
in~\cite{Fajfer:2015mia}. Most promising experimental signatures of
this model are increased branching ratios of
$B \to K^{(*)} \nu \bar \nu$ and $t \to b \tau^+ \nu$ decays. Our
results are normalized to the mass of this states to be $1$ TeV, which
is in agreement with current direct searches of CMS/ATLAS limits on
the leptoquark of the second/third
generation~\cite{Aad:2015caa,Khachatryan:2015vaa}. Further efforts on
both sides---theoretical and experimental---might help to understand
better impact and perspective of this NP candidate.

 \section*{Acknowledgements}
 We would like to thank Ilja Dor\v{s}ner and Jernej F. Kamenik for
 very useful discussions. We thank David Straub for cross-checking
and correcting the expressions for $B \to K^{(*)} \nu \bar \nu$ decay. We acknowledge
 support of the Slovenian Research Agency.

\bibliography{refs}
\end{document}